\documentclass[aps,prb,amsmath,amssymb,twocolumn,10pt,footinbib,showpacs]{revtex4-1}
\usepackage{graphicx}
\usepackage{graphics}
\usepackage{amsmath}
\usepackage{amssymb}
\usepackage{amsfonts}
\usepackage{sidecap}
\usepackage{hyperref}
\usepackage{color}
\usepackage{psfrag}

\makeatletter
\newcommand*{\rom}[1]{\expandafter\@slowromancap\romannumeral #1@}
\makeatother

\newcommand*\oline[1]{%
  \vbox{%
    \hrule height 0.5pt%                  % Line above with certain width
    \kern0.25ex%                          % Distance between line and content
    \hbox{%
      \kern-0.1em%                        % Distance between content and left side of box, negative values for lines shorter than content
      \ifmmode#1\else\ensuremath{#1}\fi%  % The content, typeset in dependence of mode
      \kern-0.1em%                        % Distance between content and left side of box, negative values for lines shorter than content
    }% end of hbox
  }% end of vbox
}

\begin{document}

\title{Transition from fractional to Majorana fermions in Rashba nanowires}
\author{Jelena Klinovaja}
\author{Peter Stano}
\author{Daniel Loss}
\affiliation{Department of Physics, University of Basel, Klingelbergstrasse 82, 4056 Basel,
Switzerland}
\date{\today}
%73.63.Fg	Nanotubes
%73.63.Nm	Quantum wires 
%74.45.+c	Proximity effects; Andreev reflection; SN and SNS junctions

\begin{abstract}
We study hybrid superconducting-semiconducting nanowires in the presence of Rashba spin-orbit interaction (SOI)  as well as helical magnetic fields.  We show that the interplay
between them leads to a competition of phases with two topological gaps closing and reopening, resulting in unexpected reentrance behavior.
Besides the topological phase with localized Majorana fermions (MFs) we find new phases characterized by fractionally charged fermion (FF) bound states of Jackiw-Rebbi type.
The system can be fully gapped by the magnetic fields alone, giving rise to FFs that transmute into MFs 
upon turning on superconductivity. We find explicit analytical solutions for MF and  FF bound states and determine the phase diagram numerically by determining the corresponding Wronskian null space. We show by renormalization group arguments that electron-electron interactions enhance the Zeeman  gaps opened by the fields.

\end{abstract}

\maketitle

{\it{Introduction.}} 
Majorana fermions~\cite{Majorana} (MF) in condensed matter systems~\cite{alicea_review_2012}, interesting from a fundamental point of view
as well as for potential applications in topological quantum computing, have attracted wide interest, both in theory 
~\cite{kitaev, fu, Nagaosa_2009, Sato, lutchyn_majorana_wire_2010, oreg_majorana_wire_2010, alicea_majoranas_2010, Akhmerov, Qi, potter_majoranas_2011, alicea_NP_2011, Brouwer_2011, Klinovaja_CNT, Chevallier} 
and experiment~\cite{mourik_signatures_2012,deng_observation_2012,das_evidence_2012}.
One of the most promising candidate systems for MFs are
semiconducting nanowires  with Rashba spin-orbit interaction (SOI) brought into proximity with a superconductor~\cite{lutchyn_majorana_wire_2010, oreg_majorana_wire_2010, alicea_majoranas_2010}.
In such hybrid systems  a topological phase with a MF at each end of the nanowire is predicted to
emerge once an applied uniform magnetic field exceeds a critical value
~\cite{Sato,lutchyn_majorana_wire_2010, oreg_majorana_wire_2010, alicea_majoranas_2010}. 
As pointed out recently~\cite{Braunecker_Jap_Klin_2009}, the Rashba SOI in such wires is equivalent to a helical Zeeman term, and thus the same
topological phase with MFs is predicted to occur in hybrid systems in the presence of  a helical field but without SOI~\cite{suhas_majorana, Flensberg_Rot_Field}.

Here, we go a decisive step further and address the question, what happens when {\it both} fields are present, an internal Rashba SOI field as well as a helical--or more generally--a spatially varying magnetic field. Quite remarkably, we discover that due to the interference between the two mechanisms the phase diagram becomes surprisingly rich, with reentrance behavior of MFs and new phases characterized by fractionally charged fermions (FF), 
analogously to Jackiw-Rebbi fermion bound states~\cite{Jackiw_Rebbi}. 
Since the system is fully gapped by the magnetic fields  at certain Rashba SOI strengths (in the absence of superconductivity), these FFs act as precursors of  MFs into which they transmute by turning on superconductivity.

The main part of this work aims at characterizing the mentioned phase diagram. For this we find explicit solutions for the various bound states, which allows us to derive analytical conditions for the  boundaries of the topological phases. We also perform an independent numerical search of the phases and present numerical results illustrating them. We show that the phases can be controlled with experimentally accessible parameters, such as the uniform field or the chemical potential. We formulate the topological  criterion as a condition local in momentum space via the kernel dimension of the Wronskian, which does not require the knowledge of the spectrum in the entire Brillouin zone.
We also address  interaction effects and show that they increase all Zeeman gaps and thereby the stability of the topological phase.

\begin{figure}[!tb]
    \centering
    \includegraphics[width=8cm]{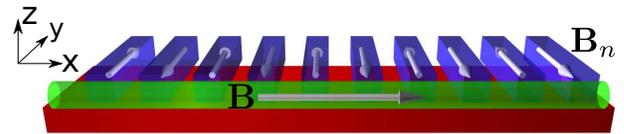}
\caption{Schematics of the hybrid semiconducting-superconducting system, consisting of a finite Rashba nanowire (green cylinder) on top of an s-wave bulk superconductor (red slab) in the presence of a uniform magnetic field $\bf B$ (grey arrow) applied along the nanowire in x-direction. Periodically arranged nanomagnets (blue bars) produce a spatially varying magnetic field ${\bf B}_n(x)$ (grey arrows). We note that ${\bf B}_n(x)$ can also be generated intrinsically e.g. by a helical hyperfine field of nuclear spins inside the nanowire~\cite{Braunecker_RG_2009}.}
\label{system}
\end{figure}

\begin{figure}[!bt]
    \centering
    \includegraphics[width=8cm]{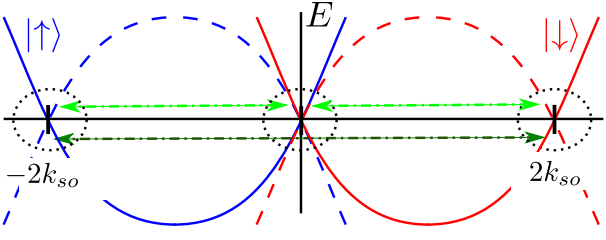}
\caption{The spectrum of Rashba nanowire consists of parabolas shifted by $\pm k_{so}$: solid and dashed lines correspond to the electron and hole spectrum, respectively. The outer circles (dotted) mark the exterior branches and the inner circle (dotted) marks the interior branches. A spatially varying magnetic field with period $4k_{so}$ (dark green arrow) couples the exterior branches. Similarly, a uniform magnetic field couples the interior branches at $k=0$ (not indicated).
A spatially varying magnetic field with period $2k_{so}$ (light green arrow) mixes exterior with interior branches.}
\label{spectrum}
\end{figure}

{\it {Model.}}
We consider a system consisting of a semiconducting nanowire with Rashba SOI in proximity with an $s$-wave bulk superconductor and in the presence of magnetic fields which contain  uniform and  spatially varying
components, see Fig.~\ref{system}.
The Rashba spin-orbit interaction is characterized by a SOI vector $\boldsymbol \alpha$ pointing along, say, the $z$-axis.  The effective continuum Hamiltonian for the nanowire is in Nambu representation given by $H_0= \frac{1}{2} \int dx\ \psi^\dagger (x) \mathcal{H}_0 \psi (x)$ with
\begin{align}
\mathcal{H}_0& =(- \hbar^2 \partial_x ^2/2m-\mu)\eta_3 - i \alpha  \eta_3 \sigma_3\partial_x   ,
\label{h_0_density}
\end{align}
where $m$ is the electron mass.
Here, $\psi =  (\Psi_\uparrow, \Psi_\downarrow, \Psi_\uparrow^\dagger, \Psi_\downarrow^\dagger)$, and $\Psi_{\sigma}^{(\dagger)}(x)$, with $\sigma=\uparrow/\downarrow$, is the annihilation (creation) operator
for a spin up/down electron at position $x$. The Pauli matrix $\sigma_i$ ($\eta_i$) acts in the spin (electron-hole) space.
The spectrum of $\mathcal{H}_0$  consists of four parabolas centered at the Rashba momentum $\pm k_{so}=\pm m\alpha/\hbar^2$, see Fig. \ref{spectrum}.
The  chemical potential $\mu$ is chosen to be zero at the crossing of the Rashba branches at $k=0$.

The uniform ($\bf B$)  and spatially varying (${\bf B}_{n}$ ) magnetic fields lead to the Zeeman term,
\begin{equation}
\mathcal{H}_{z}= g\mu_B  [{\bf B}+{\bf B}_{n}(x)]  \cdot {\boldsymbol \sigma}\ \eta_3 /2,
\end{equation}
where $g$ is the g-factor and $\mu_B$ the Bohr magneton. 
The proximity-induced superconductivity couples states of opposite momenta and  spins and is described by $\mathcal{H}_{s}= \Delta_s  \sigma_2 \eta_2$,
where the effective pairing amplitude $\Delta_s$ can be assumed to be non-negative.

From now on, we assume that the SOI energy  $ m \alpha^2/ \hbar^2$ is the largest energy scale at the Fermi level in the problem.
In this strong SOI regime, we can treat the $B$-fields and $\Delta_s$ as small perturbations. This allows us to  linearize the full Hamiltonian $\mathcal{H}_0+\mathcal{H}_{z}+\mathcal{H}_{s}$ around  $k=0$ (referred to as interior branches)  and $k=\pm 2 k_{so}$ (referred to as exterior branches), see Fig. \ref{spectrum}. 
This entails that we can use the ansatz
\begin{equation}
\Psi(x) =R_\uparrow  + L_\downarrow + L_\uparrow  e^{-2 i  k_{so} x}+ R_\downarrow  e^{2 i k_{so} x},
\label{operator}
\end{equation}
where the right mover $R_\sigma(x)$ and the left mover $L_\sigma(x)$ are slowly-varying fields.
For a uniform magnetic field alone (chosen along the $x$-axis) the full Hamiltonian becomes 
$H =\frac{1}{2}\int dx\ \widetilde\psi(x)^\dagger \mathcal{H} \widetilde\psi(x)$ with
\begin{align} 
\mathcal{H} = -i \hbar \upsilon_F  \sigma_3\tau_3\partial_x  + \Delta_z  \eta_3\sigma_1 (1+\tau_3)/2 + \Delta_s  \eta_2\sigma_2,
\label{static}
\end{align}
where the Pauli matrix $\tau_i$ acts in the interior-exterior branch space, and $\widetilde\psi=(R_\uparrow, L_\downarrow, R^\dagger_\uparrow, L^\dagger_\downarrow, L_\uparrow, R_\downarrow, L^\dagger_\uparrow, R^\dagger_\downarrow)$. The Fermi velocity is given by $\upsilon_F=\alpha/\hbar $ and the Zeeman energy by $\Delta_z=g \mu_B B/2$. 
Next, we include the spatially varying magnetic field, and assume that it has a substantial Fourier component either  at $4k_{so}$ (case ${\rm \rom{1}}$) or at $2k_{so}$ 
(case $ {\rm \rom{2}}$), leading to additional couplings between all branches of the spectrum, see Fig. \ref{spectrum}. 
We treat now the two cases in turn and will see that the interplay of Rashba and magnetic fields leads to a surprisingly reach diagram of topological phases.

{\it  Case ${\rm \rom{1}}$}.  
Here, we consider ${\bf B}_{n}(x)$ with period $4k_{so}$ and  perpendicular to  $ \boldsymbol \alpha$. 
For a  field with oscillating amplitude only, we consider two geometries, ${\bf B}_{n,x}={\hat x}B_{n}\cos (4 k_{so} x+\theta)$ and
${\bf B}_{n,y}={\hat y}B_{n}\sin (4 k_{so} x+\theta)$, with arbitrary phase shift $\theta$, while 
for a helical field we consider a field with anticlockwise rotation, $({\bf B}_{n,x}+{\bf B}_{n,y})/2$. 
(clockwise rotation does not lead to coupling).  We note that ${\bf B}_n(x)$ can also be generated intrinsically e.g. by the hyperfine field of ordered nuclear spins 
inside the nanowire~\cite{Braunecker_RG_2009}.
All geometries lead to identical results: they affect only the exterior branches (see Fig. \ref{spectrum}) and the corresponding Hamiltonian remains block-diagonal in $\tau$-space. 
The full  Hamiltonian becomes  $\mathcal{H}_{4k_{so}}=\mathcal{H}+\Delta_n  (\sigma_1  \cos \theta -  \sigma_2 \sin \theta)\eta_3 (1-\tau_3)/2$,
where $\Delta_n=g \mu_B B_n/4$. 
 The spectrum for the exterior ($l=e$) and interior ($l=i$) branches is given by
\begin{equation}
\begin{split}
E_{l}^2=(\hbar \upsilon_F k)^2&+ \Delta_s^2+\Delta_l^2+\mu^2\\
&\pm 2 \sqrt{\Delta_s^2\Delta_l^2+\mu^2 [(\hbar \upsilon_F k)^2+\Delta_l^2]},
\end{split}
\label{E_int}
\end{equation}
where  $\Delta_{e/i}=\Delta_{n/z}$. We note the equivalence of effects of a uniform field on the interior branches and of a periodic field on the exterior branches.
The  spectrum is fully gapped
except for two special  cases, $\Delta_{n/z}^2=\Delta_s^2+\mu^2$.
This suggests that there will be transitions between different non-trivial phases.

We identify these phases by the presence or absence of bound states inside the gap. For this it is most convenient to study the Wronskian  corresponding to the four decaying fundamental solutions~\cite{MF_Klinovaja}. Here, we consider a semi-infinite nanowire, with boundary at $x=0$, and assume that all decay lengths will be shorter than the system length.
For fixed parameters (including the energy $E$), we find the four decaying eigenstates of $\mathcal{H}_{4k_{so}}$  for the left and right movers. 
Using Eq.~(\ref{operator}),
we express them in the basis of the original fermionic fields $\psi$, leading to four four-spinor solutions $\Phi^j$ with $j=1,...,4$, and construct a $4 \times 4$ Wronskian matrix $W_{ij}(x)=[\Phi^j(x)]_i$. The dimension $d$ of the null space of $W(0)$ determines the system phase: $d=0$ 
corresponds to a phase with no bound states (trivial phase), 
$d=1$ at $E=0$  to  a phase with one single Majorana fermion (MF) (topological phase), $d=2$ at $E=0$ or $d=1$ at $E\neq 0$
to a phase with one localized fermion of fractional charge (FF, see below) (fermion phase). 
We refer to the trivial and fermion phases as non-topological.
Finally, the knowledge of the $W(0)$ null space allows us to construct the bound state wave functions, expressed in terms of linearly dependent combinations of $\Phi^j$ fulfilling the Dirichlet boundary condition  at $x=0$. In  App.~\ref{Supplementary} we list the analytical solutions for the MF bound states,  from which we see explicitly that these solutions are robust against any parameter variations (topologically stable) as long as the topological gap remains open.

\begin{figure}[tb]
    \centering
    \includegraphics[width=8.5cm]{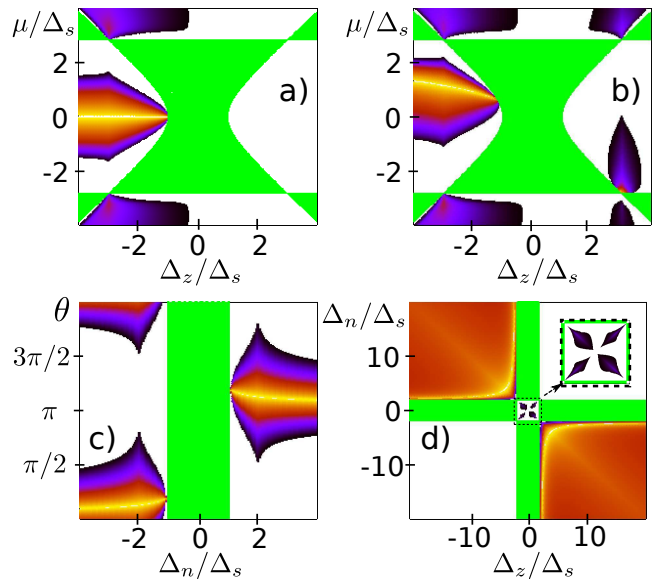}
\caption{Phase diagram for case I supporting three phases: the trivial phase with no bound states (white area), the topological phase with one MF (green area), and the fermion phase with two FFs (colored area). The color in the fermion phase encodes the ratio of the fermion energy to the system gap, which varies from zero (bright yellow) to one (black).
Note that the central  region of the topological phase corresponds to case IA, while
the four corner regions to case IB. Here, (a) $\Delta_n/\Delta_s=3$, $\theta=0$, while (b) $\Delta_n/\Delta_s=3$, $\theta=\pi/4$. Note that $\Delta_z<0$ and $\theta$ corresponds to $\Delta_z>0$ and $\theta+\pi$.
By comparing (a) with (b) and by calculating the dependence on $\theta$ [(c) $\mu/\Delta_s=0.4$, $\Delta_z/\Delta_s=2$], we note that the topological phase is insensitive to $\theta$ in contast to the fermion phase. The position of a zero-energy FF depends not only on $\theta$ and $\mu$ [see (a)-(c)] but also on $\Delta_z$ and $\Delta_n$ [(d) $\mu/\Delta_s=1.5$, $\theta=\pi/4$] in agreement with Eq. (\ref{FphaseCondition4kso}).
}
\label{phase_diagram}
\end{figure}

For case I, we find that the system is in the topological phase if one of the following inequalities is satisfied, 
\begin{eqnarray}
&{\rm (\rom{1}A)} \ \ \ \  \Re \sqrt{\Delta_n^2-\mu^2}<\Delta_s<\Re \sqrt{\Delta_z^2-\mu^2}\ \ \ \ \
\label{MajoranaCondition4ksoI}\\
&{\rm (\rom{1}B)} \ \ \ \  \Re\sqrt{\Delta_z^2-\mu^2}<\Delta_s<\Re \sqrt{\Delta_n^2-\mu^2},\ \ \ \ 
\label{MajoranaCondition4ksoII}
\end{eqnarray}
with the corresponding MF wave functions given in App.~\ref{Supplementary}.
Case IA goes into IB upon interchange $\Delta_z \leftrightarrow \Delta_n$.
As anticipated after Eq.~(\ref{E_int}), the  boundaries of the topological phase correspond to the system being gapless. In the absence of ${\bf B}_n$, there is only one topological gap, which arises from the interior branches \cite{lutchyn_majorana_wire_2010, oreg_majorana_wire_2010}. In this case, only condition IA can be satisfied, and a MF emerges when the uniform $B$-field exceeds a critical value. However, in the presence of ${\bf B}_n$, the exterior gap is also topological. As shown in Fig.~\ref{phase_diagram}, the interplay between the two gaps leads to a rich phase diagram with reentrance behaviour.  For instance, if $|\Delta_{l}| > |\mu|$ and $\Delta_s=0$,  the system is  in the non-topological phase but still fully  gapped by the magnetic fields. With increasing  $\Delta_s$, first the exterior (interior)  gap closes and reopens, bringing the system into the topological phase. Then, upon further increase of $\Delta_s$, the interior (exterior) gap closes and reopens, bringing the system back into the non-topological phase. 

We note that case IB allows the presence of a MF in weaker uniform magnetic fields, see Fig. \ref{phase_diagram}.
If the nanomagnets generating ${\bf B}_n$ can be arranged such that the field penetration into the bulk-superconductor is minimized, as illustrated in Fig.~\ref{system},
much stronger oscillating  than uniform fields can be applied, opening up the possibility to generate MFs in systems with small $g$-factors.

The system is in the fermion phase, if 
\begin{equation}
 \begin{split}
&\theta = \pi+\phi_n+\phi_z,\, {\rm and} \\
&\Delta_s<{\rm min}\{\Re \sqrt{\Delta_z^2-\mu^2}, \Re \sqrt{\Delta_n^2-\mu^2} \},  
\end{split}
\label{FphaseCondition4kso}
\end{equation}
where the phases $\phi_{z,n}$ are defined by $e^{i\phi_{z,n}}=\left(\sqrt{\Delta_{z,n}^2-\mu^2} + i \mu\right)/\Delta_{z,n}$. The corresponding wave functions are listed in App.~\ref{Supplementary}.
In this regime, two MFs (both localized at $x=0$) fuse to one fermion bound state.
Such bound state fermions are known to have fractional charge $e/2$~\cite{FracCharge_Kivelson, FracCharge_Bell,FracCharge_Chamon}, 
as first discovered in the Jackiw-Rebbi model~\cite{Jackiw_Rebbi,CDW}. 
 If neither of the inequalities~(\ref{MajoranaCondition4ksoI})-(\ref{FphaseCondition4kso}) is satisfied, the system is in the trivial phase without any bound state at zero energy. 

Detuning from the conditions in Eq.~(\ref{FphaseCondition4kso}), the two zero-energy solutions are usually split, becoming a fermion-antifermion pair at energies $\pm E$. Importantly, FFs do not require the presence of superconductivity. For example, if $\Delta_s=0$ and $\mu=0$, the two bound states have energy
\begin{equation}
E_{FF} = \pm \frac{\Delta_z \Delta_n \sin\theta}{\sqrt{\Delta_z^2+\Delta_n^2-2\Delta_z\Delta_n\cos\theta}}.
\label{FF_energy}
\end{equation}
We note that the splitting vanishes at $\theta=n\pi$, $n$ integer, due to the chiral symmetry of $\mathcal{H}_{4k_{so}}$ at these special values~\cite{CDW}.
In contrast,  the MF remains at zero energy for all values of $\theta$, which is a direct manifestation of the stability of the bound state within the topological phase (despite the fact that the MF wave function depends on $\theta$, see App.~\ref{Supplementary}).

To determine the full phase diagram we have performed a systematic numerical search  for all bound state solutions with energies 
inside the gap and determined the null space of the Wronskian. The results are plotted in Fig. \ref{phase_diagram}.
The bright yellow lines inside the colored area in Fig. \ref{phase_diagram} correspond to  zero-energy FFs  satisfying Eq.~(\ref{FphaseCondition4kso}). 
At the point where the lines  touch the topological phase (shown in green), the gap
closes and reopens and {\it the zero-energy} FF {\it transmutes into a} MF.
In the fermion phase away from the zero-energy line  the two solutions split (the bigger the splitting the darker the color), until they finally reach the gap (black boundaries) and disappear. The fermion phase exists only for certain values of the phase shift $\theta$, in contrast to the topological phase, which, again, is not sensitive to $\theta$, see Fig. \ref{phase_diagram}c. Moreover, the fermion phase is also sensitive to the relative orientation of $\bf B$ and ${\bf B}_n$, see Fig. \ref{phase_diagram}d.  In the same panel, we see that outwards regions are more suitable for a fractional charge observation than the central region, where energies of the bound states are very close to the gap edge.

{\it  Case ${\rm \rom{2a}}$}.  
We now shortly comment on two addititonal geometries with a spatially varying magnetic field with a period $2k_{so}$. To keep the following discussion concise, we set $\mu=0$ and state the results for $E=0$ bound states only.  We begin with a field perpendicular to the SOI vector $\boldsymbol \alpha$ and given by ${\bf B}_{n,x}={\hat x}B_{n}\cos (2 k_{so} x)$ or
${\bf B}_{n,y}={\hat y}B_{n}\sin (2 k_{so} x)$ (an oscillating field) or $({\bf B}_{n,x}+{\bf B}_{n,y})/2$ (a helical field). Such a field mixes the exterior and interior branches, see Fig. \ref{spectrum}. The corresponding Hamiltonian is  $\mathcal{H}_{2k_{so}}^\perp=\mathcal{H}+\Delta_n\eta_3\sigma_1\tau_1$. The spectrum is given by
\begin{align}
E_\perp^2&=(\hbar \upsilon_F k)^2+[\Delta_n^2+\Delta_z^2/4]+(\Delta_s\pm \Delta_z/2 )^2\\
&\pm2\sqrt{ (\hbar \upsilon_F k)^2 \Delta_n^2+[\Delta_n^2+\Delta_z^2/4](\Delta_s\pm \Delta_z/2 )^2} \nonumber.
\end{align}

Repeating the procedure used above, we derive the condition for the topological phase as
\begin{align}
\Delta_z>\Delta_s\left|1-\Delta_n^2/\Delta_s^2\right|. \label{2kso_top}
\end{align}
Again, the phase boundary to the topological phase corresponds to the parameters at which the system is gapless, i.e. $\Delta_z=\Delta_s\left|1-\Delta_n^2/\Delta_s^2\right|$. 
We note that in the presence of a spatially periodic magnetic field MFs may emerge at substantially weaker uniform magnetic fields.
The fermion phase occurs if 
\begin{equation}
 |\Delta_n|>\Delta_s \ {\rm{and}} \ \ \Delta_z<\Delta_s\left(\Delta_n^2/\Delta_s^2-1\right).
\end{equation}
The rest of the parameter space corresponds to the non-topological phase.
The corresponding wave functions for MFs and FFs are given in App.~\ref{Supplementary}.

{\it  Case ${\rm \rom{2}b}$}.  
Finally, we comment on an oscillating field aligned with the SOI vector $\boldsymbol{\alpha}$ and given by ${\bf B}_{n} =  \hat{z} B_n\cos (2 k_{so} x + \theta)$.
 This field couples the interior and exterior branches (see Fig. \ref{spectrum}).
The corresponding Hamiltonian is  $\mathcal{H}_{2k_{so}}^\parallel=\mathcal{H}+\eta_3\sigma_3\tau_1\cos\theta-\tau_2\sin\theta$,
with the spectrum given by 
\begin{align}
E_\parallel^2&=(\hbar \upsilon_F k)^2+\left[\sqrt{\Delta_n^2+\frac{\Delta_z^2}{4}}\pm\left(\frac{\Delta_z}{2}\pm \Delta_s\right)\right]^2.
\end{align}
The topological phase is determined, again, by Eq. (\ref{2kso_top}) and
 the corresponding MF wave functions are given in App.~\ref{Supplementary}. We note once more that MFs can be observed in weaker uniform magnetic fields.
Interestingly, in this configuration the fermion phase is absent,  demonstrating the sensitivity of the FFs to the $B$-field orientation.

{\it Interactions.} Electron interactions play an important role in one-dimensional systems~\cite{Giamarchi} and, in particular, for MFs~\cite{suhas_majorana,stoudenmire}. E.g., the interior gap opened by a uniform magnetic field is strongly enhanced by interactions ~\cite{Braunecker_Jap_Klin_2009}; and so we expect the same renormalization to occur here for both gaps. This is indeed the case, as we show next.
For this, we perform a renormalization group analysis for both the uniform and the periodic field.
Following Ref. \cite{Braunecker_RG_2009} we arrive at the effective Hamiltonian 
$H=\sum_l \int \frac{d x}{2 \pi} H_l (x)$ in terms of conjugate boson fields, $\phi_{e,i}$ and $\theta_{e,i}$, with
\begin{align}
H_l= \upsilon [(\nabla \phi_l )^2+ (\nabla \theta_l )^2]+ \frac{\Delta_l}{a} \cos (2\sqrt{ K} \phi_l),
\end{align}
where we have suppressed  quadratic off-diagonal  terms being less relevant compared to the cosine terms. The index $l=e/i$ denotes the exterior/interior branch, $a$  the lattice constant, $K^2 = ( \upsilon_s/K_s+\upsilon_c K_c)/(\upsilon_c/K_c+\upsilon_s K_s)$ and $\upsilon=\sqrt{\left(\upsilon_c/K_c+\upsilon_s K_s\right)\left(\upsilon_s/K_s+\upsilon_c K_c\right)}/2$. Here, $\upsilon_{c,s}$ are the charge (c) and spin (s)  velocities and $K_{c,s}$  the corresponding Luttinger liquid parameters~\cite{Giamarchi}. The gaps $\Delta_l$ opened by magnetic fields
are renormalized upwards by interactions and given by $\widetilde{\Delta}_l = \Delta_l (\pi \hbar \upsilon_F/\Delta_l a)^{(1-K)/(2-K)}$. For GaAs (InAs) nanowires~\cite{Braunecker_RG_2009,suhas_majorana}, we estimate an increase by about a factor of 2 (4).
 The enhanced Zeeman gaps allows the use, again, of materials with lower $g$-factors. 
 
{\it Conclusions.} The interplay between spatially varying magnetic fields and Rashba SOI in a hybrid nanowire system leads to a rich phase diagram with reentrance behavior and with fractionally charged fermions
that get transmuted into Majorana fermions at the reopening of the topological gap. 
%The presence of a spatially varying field and the effects of interactions opens up the possibility to use materials with lower g-factors.

\acknowledgments
We acknowledge useful discussions with Claudio Chamon.
This work is supported by the Swiss NSF, NCCR Nanoscience, and NCCR QSIT.

\appendix

\section{Analytical solutions \label{Supplementary}}

Here we present the analytical solutions of the Majorana fermion (MF) and fractional fermion (FF) wave functions for different regimes discussed in the main text. Every MF wave function has the four-spinor form $\varPhi_{MF} (x)=(f(x), g(x), f^*(x), g^*(x))$
with normalization condition $\int dx\ |\varPhi_{MF}(x)|^2=2$ (below we omit the normalization factors). For zero energy $E=0$, the functions $f(x)$ and $g(x)$ are listed in Table~\ref{wavefunctions_supp}, together
with the regime of validity corresponding to the cases I and II defined in the main text.

We use the notations $e^{i\phi_z}=(\sqrt{\Delta_z^2-\mu^2} + i \mu)/\Delta_z$, $ e^{i\phi_n}=(\sqrt{\Delta_n^2-\mu^2} + i \mu)/\Delta_n$, and $ \phi_{\pm}= (\theta\pm\phi_n-\phi_z)/2$. 
In row IIa-FF of  Table~\ref{wavefunctions_supp}, only one MF wave function is given, a second one needs to be added from row IIa-MF, depending on whether $(\Delta_s-\Delta_z/2)^2>\Delta_n^2$ or
$(\Delta_s-\Delta_z/2)^2<\Delta_n^2$.

If $\Delta_s=0$ and $\mu=0$, the fractional fermions have energies given be Eq.~\ref{FF_energy}
% \begin{equation}
% E_{FF} = \Delta_z \Delta_n\sin\theta/\sqrt{\Delta_z^2+\Delta_n^2-2\Delta_z\Delta_n\cos\theta},
% \end{equation}
and the corresponding wave function is given by
\begin{equation}
\varPhi_F=\begin{pmatrix}
           \varPhi_\uparrow \\ \varPhi_\downarrow
          \end{pmatrix}
= \begin{pmatrix}
   e^{i \phi^{F}} \\
1
  \end{pmatrix}
 \left(e^{2i k_{so} x{- x/\xi_2^F}} - e^{-x/\xi_1^F }\right),
\end{equation}
where $\cos\phi^{F}=E_{FF}/\Delta_z$,  $\xi_1^F=\alpha/\sqrt{\Delta_z^2-E_{FF}^2}$, and $\xi_2^F=\alpha/\sqrt{\Delta_n^2-E_{FF}^2}$. We note that this fermion of  non-negative energy has an anti-fermion partner of non-positive energy.

\begin{table*}[hbt]
\caption{Wave functions of MFs and FFs for different regimes at zero energy.
}
\begin{tabular}{l|l}

\hline\hline\\[-10pt]
&$\sqrt{\Delta_n^2-\mu^2}<\Delta_s<\sqrt{\Delta_z^2-\mu^2}$, \ \ $|\Delta_{z,n}| > |\mu|$
\\[5pt]
${\rm (\rom{1}A)}$
& 
$f= i g^*=ie^{- x/\xi_1^{(e)}+i (2 k_{so}x+\phi_++\phi_z/2) } \cos \phi_-  +  e^{-x/\xi_3^{(e)}+i (2k_{so}x+\phi_-+\phi_z/2)}\sin \phi_+  -i  e^{-x/\xi_2^{(i)}+i\phi_z/2}\cos \phi_n$ \\[5pt]
MF& $\xi_{1}^{(e)} = \alpha/(\Delta_s - \sqrt{\Delta_n^2-\mu^2})$,\ $\xi_{3}^{(e)} = \alpha/(\Delta_s + \sqrt{\Delta_n^2-\mu^2})$,\ $\xi_{2}^{(i)}=\alpha/(\sqrt{\Delta_z^2-\mu^2}-\Delta_s)$
\\[5pt] \cline{2-2} \\[-10pt]
& $0<\Delta_s<\sqrt{\Delta_z^2-\mu^2}$, $|\Delta_{n}| = |\mu|$,\  $|\Delta_{z}| > |\mu|$\\[5pt]
& $f= 
e^{-\Delta_s x/\alpha+i2k_{so}x}(-i e^{i \phi_z/2}+ (2 x\mu/\alpha)e^{-i (\theta/2+\pi/4)}\sin[(\theta-\phi_z)/2+\pi/4]) +i e^{- x/\xi_2^{(i)}+i \phi_z/2}$\\[5pt]
& $g= 
e^{-\Delta_s x/\alpha-i2k_{so}x}(- e^{-i \phi_z/2}+ (2 x\mu/\alpha)e^{i (\theta/2+3\pi/4)}\sin[(\theta-\phi_z)/2+\pi/4]) + e^{- x/\xi_2^{(i)}-i \phi_z/2}$
\\[5pt] \cline{2-2} \\[-10pt]
& $0<\Delta_s<\sqrt{\Delta_z^2-\mu^2}$, $|\Delta_{n}| < |\mu|$,\  $|\Delta_{z}| > |\mu|$; \ \ \ \ \ \ \  $k_m=\sqrt{\mu^2-\Delta_n^2}/\alpha$\\[5pt]
& $f= i g^*=i e^{- x/\xi_2^{(i)}+i \phi_z/2} -i  e^{-\Delta_s x/\alpha + i( 2 k_{so} x+ \phi_z/2)}\left(\cos (k_m x) + \sin (k_m x)\left[ i \mu + \Delta_n e^{i(\theta-\phi_z)}\right]/\sqrt{\mu^2-\Delta_n^2}\right)$ 
\\[5pt]
\hline\\[-10pt]
&$ \sqrt{\Delta_z^2-\mu^2}<\Delta_s<\sqrt{\Delta_n^2-\mu^2}$, \ \ $|\Delta_{z,n}| > |\mu|$ \\[5pt]
${\rm (\rom{1}B)}$
& 
$f=-i g^*= - i  e^{- x/\xi_1^{(i)}-i\phi_z/2} \cos \phi_- +e^{-x/\xi_3^{(i)} +i\phi_z/2}\sin [(\theta-\phi_n+\phi_z)/2]+i e^{-x/\xi_2^{(e)}+i(2k_{so}x+\phi_-+\phi_z/2)} \cos \phi_z $ \\[5pt]
MF& $\xi_{1}^{(i)}=\alpha/(\Delta_s - \sqrt{\Delta_z^2-\mu^2})$, $\xi_{3}^{(i)}=\alpha/(\Delta_s + \sqrt{\Delta_z^2-\mu^2})$, $\xi_{2}^{(e)} = \alpha/(\sqrt{\Delta_n^2-\mu^2}-\Delta_s)$
\\[5pt] \cline{2-2} \\[-10pt]
& $\Delta_s<\sqrt{\Delta_n^2-\mu^2}$,\  $\Delta_z=|\mu|$,\ $|\Delta_{n}| > |\mu|$\\[5pt]
& $f= e^{-\Delta_s x/\alpha}\left( i e^{i(\theta-\phi_n)/2}-(2 x\mu/\alpha) e^{i \pi/4}\sin\left[(\phi_n-\theta)/2-\pi/4\right]\right) -i e^{- x/\xi_2^{(e)}+i(4k_{so}x+\theta-\phi_n)/2}$\\[5pt]
& $g= e^{-\Delta_s x/\alpha}\left(-e^{-i(\theta-\phi_n)/2} +(2 x\mu/\alpha)e^{i \pi/4}\sin\left[(\phi_n-\theta)/2-\pi/4\right]\right)+e^{-x/\xi_2^{(e)}+i(4k_{so}x-\theta+\phi_n)/2}$\\[5pt] \cline{2-2} \\[-10pt]
& $\Delta_s<\sqrt{\Delta_n^2-\mu^2}$,  $|\Delta_z|<|\mu|$, $|\Delta_{n}| > |\mu|$; \ \ \ \ \ \ \ \ \ \  $k_m =\sqrt{\mu^2-\Delta_z^2}/\alpha $ \\[5pt]
& $f=-i g^*=e^{i(\theta-\phi_n)/2} \left[ie^{-x/\xi_2^{(e)}+i 2 k_{so}x}-i e^{-\Delta_s x/\alpha} \left(\cos(k_m x)+\sin(k_m x)\left[\Delta_z e^{i (\phi_n-\theta)}-i \mu\right]  /\sqrt{\mu^2-\Delta_z^2} \right)\right]$ \\[5pt]
\hline\\[-10pt]
(I) &$\theta = \pi+\phi_n+\phi_z$, $0\leq\Delta_s<{\rm min}\{\Re \sqrt{\Delta_z^2-\mu^2}, \Re \sqrt{\Delta_n^2-\mu^2}\}$, $E_{FF}=0$
\\[5pt]
FF&$f_1 =i g^*_1= i e^{i\phi/2} \left(e^{{2ik_{so}x}-x/\xi_3^{(e)}} - e^{-x/\xi_2^{(i)}} \right)$, $f_2=-i g^*_2=e^{i\phi/2}\left( e^{{2ik_{so}x}-x/\xi_2^{(e)}} -e^{- x/\xi_3^{(i)}}\right)$ \\[5pt]
 \hline\hline\\[-10pt]
&$\Delta_z>\Delta_s\left|1-\Delta_n^2/\Delta_s^2\right|$ and $(\Delta_s-\Delta_z/2)^2>\Delta_n^2$; \ \ \ \ \  $\xi_{\pm}=\alpha/\left(\Delta_z/2\pm\sqrt{(\Delta_s-\Delta_z/2)^2 - \Delta_n^2}\right)$ \\[5pt]
${\rm (\rom{2}a)}$
&
$f=ig^*= i e^{2ik_{so}x}  \left(e^{-x/\xi_-} \Delta_n + e^{- x/\xi_+} A\right)-i e^{- x/\xi_+} \Delta_n -i e^{- x/\xi_-} A $, $A= \alpha/ \xi_--\Delta_s$ \\[5pt]
\cline{2-2} \\[-10pt]
MF&$\Delta_z>\Delta_s\left|1-\Delta_n^2/\Delta_s^2\right|$  and $(\Delta_s-\Delta_z/2)^2\leq\Delta_n^2$; \ \ \ \ $\cos \phi = (\Delta_z/2-\Delta_s)/\Delta_n$,  $k_{m}=\sqrt{\Delta_n^2-(\Delta_s-\Delta_z/2)^2 }/\alpha$\\[5pt]
& $f=i g^*=\left[ i e^{2ik_{so}x}  \cos (k_{m} x +\phi/2)-i  \cos (k_{m} x -\phi/2) \right] e^{-\Delta_z x/2\alpha}$\\[5pt]
\hline  \\[-10pt]
 ${\rm (\rom{2}a)}$
&$|\Delta_n|>\Delta_s$,   $\Delta_z<\Delta_s\left(\Delta_n^2/\Delta_s^2-1\right)$, and $(\Delta_s+\Delta_z/2)^2>\Delta_n^2$; \ \ \ \ \ $\widetilde{\xi}_{\pm}=\alpha/\left(\Delta_z/2\pm\sqrt{(\Delta_s+\Delta_z/2)^2 - \Delta_n^2}\right)$ \\[5pt]
FF
&
$f=-ig^*=e^{2ik_{so}x} \left(e^{- x/\widetilde{\xi}_{-}} \Delta_n + e^{-x/\widetilde{\xi}_{+}} B\right)-e^{- x/\widetilde{\xi}_{+}} \Delta_n - e^{- x/\widetilde{\xi}_{-}} B $,\ \ $B=\alpha/ \widetilde{\xi}_{-}+\Delta_s  $ \\[5pt]
\cline{2-2} \\[-10pt]
& $|\Delta_n|>\Delta_s$,   $\Delta_z<\Delta_s\left(\Delta_n^2/\Delta_s^2-1\right)$, and $(\Delta_s+\Delta_z/2)^2\leq\Delta_n^2$;\ \   $\cos \phi = (\Delta_z/2+\Delta_s)/\Delta_n$,   $k_{m}=\sqrt{\Delta_n^2-(\Delta_s+\Delta_z/2)^2}$ \\[5pt]
& $f=-ig^*=\left[   \cos (k_{m} x -\phi/2) 
-   e^{i2k_{so}x}  \cos (k_{m} x +\phi/2)\right] e^{-\Delta_z x/2\alpha {-i\phi/2} }$\\[5pt]
\hline\hline \\[-10pt]
&  $\Delta_z>\Delta_s\left|\Delta_n^2/\Delta_s^2-1\right|$\\[5pt]
${\rm (\rom{2}b)}$
&$f=\Delta_n[e^{-x/\xi_2}\cos\theta-ie^{- x/\xi_4}\sin\theta- e^{{- x/\xi_3} +i(2k_{so} x-\theta)}]$\\[5pt]
MF&\ \ \ \ \ \ \ \ \ \ \ \ \ \ \ \ \ \ \ \ \ \ \ \ \ \ $+i [\Delta_{z}/2+\sqrt{ \Delta_n^2+\Delta_z^2/4}](e^{i(2k_{so} x-\theta)}  (e^{- x/\xi_2}\cos\theta+ie^{-x/\xi_4}\sin\theta) - e^{- x/\xi_3})$\\[5pt]
&$g=-i\Delta_n [ e^{- x/\xi_2}\cos \theta+ i e^{- x/\xi_4}\sin \theta-  e^{{- x/\xi_3}-i(2k_{so} x-\theta)}]$\\[5pt]
    &\ \ \ \ \ \ \ \ \ \ \ \ \ \ \ \ \ \ \ \ \ \ \ \ \ \ \ \ \  $+[\Delta_{z}/2+\sqrt{ \Delta_n^2+\Delta_z^2/4}] (e^{-i(2k_{so} x-\theta)} (e^{- x/\xi_2} \cos\theta-i e^{-x/\xi_4 }\sin\theta )-e^{- x/\xi_3})$\\[5pt]
& $\xi_{2}=\alpha/\left(\Delta_z/2 + \Delta_s-\sqrt{\Delta_n^2+\Delta_z^2/4} \right)$, $\xi_3=\alpha/\left(\Delta_z/2 - \Delta_s+\sqrt{\Delta_n^2+\Delta_z^2/4} \right)$,  $\xi_4=\alpha/\left(\Delta_s+\sqrt{\Delta_n^2+\Delta_z^2/4}-\Delta_z/2  \right)$
\\[3pt]
\hline\hline 
\end{tabular}
\label{wavefunctions_supp}
\end{table*}

\end{document}